\documentclass{JHEP3}
\usepackage{amsmath}
\usepackage{amsfonts,amssymb}
\usepackage{graphics,epsfig}
\title{Spectral action, Weyl anomaly
and the Higgs-Dilaton potential}

\author{A.A.~Andrianov$^{abc}$, M.A. ~Kurkov$^{de}$, Fedele Lizzi\,$^{bcde}$~\\
        $^{a}$ V.A. Fock Department of Theoretical Physics, Sankt-Petersburg State University,
198504 St. Petersburg, Russia\\
$^{b}$\it High Energy Physics Group, Dept. Estructura i Constituents
de la Mat\`eria, \\Universitat de Barcelona, Diagonal 647, 08028
Barcelona, Catalonia, Spain \\
$^{c}$
Institut de Ci\`encies del Cosmos, UB, Barcelona\\
 $^{d}$ Dipartimento di Scienze Fisiche, Universit\`{a} di
Napoli {\sl Federico II}~\\ $^e$ INFN, Sezione di Napoli\\
Monte S.~Angelo, Via Cintia, 80126 Napoli, Italy\\
        E-mail: \email{andrianov@icc.ub.edu,  katok86@mail.ru, fedele.lizzi@na.infn.it\\
        }}

\def\appendix#1{\addtocounter{section}{1}\setcounter{equation}{0}
\renewcommand{\thesection}{\Alph{section}}
\section*{
\thesection~
{#1}}
\addcontentsline{toc}{section}{
\thesection\ \ \ #1} }

\newcommand{\no}{\nonumber\\}
\newcommand{\e}{{\rm e}}
\newcommand{\ii}{{\rm i}}
\newcommand{\dd}{{\rm d}}
\newcommand{\eqn}[1]{(\ref{#1})}
\newcommand{\be}{\begin{equation}}
\newcommand{\ee}{\end{equation}}
\newcommand{\beq}{\begin{equation}}
\newcommand{\eeq}{\end{equation}}
\newcommand{\bea}{\begin{eqnarray}}
\newcommand{\eea}{\end{eqnarray}}
\newcommand{\ba}{\begin{eqnarray}}
\newcommand{\ea}{\end{eqnarray}}

\newcommand{\la}[1]{\label{#1}}

\def\bra#1{\left\langle #1\right|}
\def\ket#1{\left| #1\right\rangle}

\def\ketbra#1#2{\left| #1\right\rangle\left\langle #2\right|}

\newcommand{\tr}[1]{\:{\rm tr}\,#1}
\newcommand{\Tr}[1]{\:{\rm Tr}\,#1}
\def\one{\mbox{1 \kern-.59em {\rm l}}}
\newcommand{\del}{\partial}

\newcommand{\complex}{{\mathbb C}} 
\abstract{ We show how the bosonic spectral action emerges from
the fermionic action by the renormalization group flow in the
presence of a dilaton and the Weyl anomaly. The induced action
comes out to be basically the Chamseddine-Connes spectral action
introduced in the context of noncommutative geometry. The entire
spectral action describes gauge and Higgs fields coupled with
gravity. We then consider the effective potential and show, that
it has the desired features of a broken and an unbroken phase,
with the roll down.}  \preprint{ICCUB-11-147\\DSF/7/2011}

\begin{document}

\tableofcontents

\section{Introduction}
In this note we will show the intimate relationships between Weyl
anomalies, the dilaton and the Higgs field in the framework of
spectral physics. The framework is the expression of a field theory
in terms of the spectral properties of a (generalized) Dirac
operator. In this respect this work can be seen in the framework of
the noncommutative geometry approach to the standard model of Connes
and collaborators~\cite{Connesbook, ConnesLott, SpectralAction,
AC2M2}, as well as of Sakharov induced gravity~\cite{Sakharov} (for
a modern review see~\cite{Visser}).

We start with a generic action for a chiral theory of fermions
coupled to gauge fields and gravity. The considerations here apply
to the standard model, but we will not need the details of the
particular theory under consideration. It is known, and this is the
essence of the noncommutative geometry approach to the standard
model, that the theory is described by a fermionic action and a
bosonic action, both of which can be expressed in terms of the
spectrum of the Dirac operator. In~\cite{anlizzi} two of us have
shown that if one starts from the classic fermionic action and
proceeds to quantize the theory with a regularization based on the
spectrum, an anomaly appears. it is possible that the full quantum
theory is still invariant by correcting the path integral measure.
This is tantamount to the addition of a term to the action, which
renders the bosonic background interacting to the dilaton field. The
main result of that paper is that this term is a modification of the
bosonic spectral action~\cite{SpectralAction}. In this case the
theory is still invariant.

In this paper we have a shift of the point of view. We still
consider the theory to be regularized in the presence of a cutoff
scale, but we consider this scale to have a physical meaning, that
of the breaking of Weyl invariance. We then consider the flow of the
theory at a renormalization scale, which is not necessarily the
scale which breaks the invariance. The theory has a dilaton, and the
Higgs field.

The dilaton may involve a collective scalar mode of all fermions
accumulated in a {Weyl-noninvariant} dilaton action. Accordingly the
spectral action arises as a part of the fermion effective action
divided into the Weyl non-invariant and Weyl invariant parts.

We calculate the dilaton effective potential  and we discuss how it
relates to the transition from the radiation phase with zero vacuum
expectation value of Higgs fields and massless particles to the
electroweak broken phase via condensation of Higgs fields. The
collective field of dilaton can provide the above mentioned phase
transition with EW symmetry breaking during the evolution of the
universe.

The next five
Sections of the paper will present the general framework, namely the
Weyl invariance of fermions in a fixed background described via the
(generalized) Dirac operator in Section 2, the connections with
noncommutative geometry, the Weyl invariance properties, the
spectral action and the bosonic action in the the following
Sections.  These sections mostly follow reference~\cite{anlizzi},
although the point of view presented in Sect.~5 is different, and in
particular show two possible ways to obtain the spectral action,
which is briefly introduced in Sect.~6. The cosmological
implications are discussed in Sec.~7. This material has not been
previously published, but parts of it have been presented in a
conference~\cite{corfuproc}. A final Section contains the
conclusions.

\section{Fermions in a Fixed Background}
Our starting point is a theory in which we have some matter fields,
represented by fermions transforming under some (reducible)
representation a gauge group, such as the standard model group
$SU(3)\times SU(2)\times U(1)$. We need not specify the group for
the moment. The fermions will be spinors belonging to some Hilbert
space $\cal H$ which we assume to be ``chiral'', i.e.\ split into a
left and a right spaces:
\be
{\cal H}={\cal H}_L\oplus{\cal H}_R
\ee
A generic matter field will therefore be a spinor
\be
\Psi=\begin{pmatrix}\Psi_L\\
\Psi_R\end{pmatrix} \label{LRspinors}
\ee
and in this representation the chirality operator, which we call
$\gamma$, is a two by two diagonal matrix with plus and minus one
eigenvalues. The two components are spinors themselves and we are
not indicating the gauge indices, nor the flavor indices. We will
assume that the fermions come in a number of identical generations,
distinguished only by the masses (or more precisely their Yukawa
coupling).

The dynamics of the fermions is given by coupling them to a gauge
and gravitational background. This coupling is performed by a
classical action, which we schematically write as:
\be
S_F=\bra{\Psi}D\ket{\Psi} \label{fermionicaction}
\ee
The operator $D$~\cite{SpectralAction} is a $2\times 2$ matrix acting on
spinors of the kind~\eqn{LRspinors}
\be
D=\left(\begin{array}{cc} i\gamma^\mu D_\mu + {\mathbb A} & \gamma_5 S\\
 \gamma_5 S^\dagger & i\gamma^\mu D_\mu + {\mathbb
 A}\end{array}\right) \label{dirac22}
\ee
where \small
\be
D_\mu=\del_\mu+\omega_\mu,
\ee
the quantity $\omega_\mu$ is the spin connection, $\mathbb A$
contains all gauge fields of the theory and $S$ contains the
information about Higgs field, Yukawa couplings, mixings and all
terms which couple the left and right part of the spinors. The
gravitational background is in general nontrivial, and the metric is
encoded in the anticommutator of the $\gamma$'s:
$\{\gamma^\mu,\gamma^\nu\}=2 g^{\mu\nu}$.

%


The quantity $\mathbb A$ represents instead a fixed gauge
background, and the interaction of the spinors with it. We emphasize
that at this stage we are just describing the classical dynamics of
fermions in a fixed background. We are deliberately vague as to the
detail of the model at this stage, not discussing important elements
of the theory, like chirality or charge conjugation. The scheme
presented here is largely independent on the details of the model.
In particular it applies to the standard model, especially in the
approach based on noncommutative geometry introduced by Connes,
which we briefly describe in Sect.~\ref{Connesspectral}.

\section{Weyl invariance and the Fermionic Action}
The fermionic action~\eqn{fermionicaction} in invariant under the
transformation
\bea
\ket{\Psi
}&\to&\e^{\frac\phi2}\ket{\Psi}\nonumber\\
D&\to& \e^{-\frac\phi2}D\e^{-\frac\phi2} \label{weyloperatorform}
\eea
where the operator $\phi$ is a function of the (operator) $x$, or in
a simpler case a constant. The action~\eqn{fermionicaction} can be
expressed in coordinates as\footnote{We use the following normalization of eigenstates $|x\rangle$
of the coordinate operator: $\langle x |y\rangle = \delta(x-y)$, that corresponds to $\int d^4 x |x\rangle\langle x | = 1$ (without $\sqrt{|g|}$), and consequently such a normalized
$|x\rangle$ doesn't transform under the Weyl transformation $g_{\mu\nu}\rightarrow e^{2\phi}g_{\mu\nu}$.}
\be
S_F=\int d^4x \sqrt{|g|} \psi(x)^\dagger D_x \psi(x);\quad (|g|)^{1/4}\psi(x) = \langle{x}\ket{\Psi}
\ee
where we introduced the subscript $x$ on $D$ to stress the fact that
it is an operator acting on the $x$ coordinate. The
transformation~\eqn{weyloperatorform} can be seen as a (generalized)
Weyl transformation\footnote{One has to pay attention to the measure
in checking transformations and Hermiticity of the operators.}:
\bea
g_{\mu\nu}(x)&\to&\e^{2\phi(x)} g_{\mu\nu}(x) \nonumber\\
\psi(x)&\to& \e^{-\frac32\phi(x)} \psi(x)\nonumber\\
(D_x\psi(x))&\to& \e^{-\frac52\phi(x)}(D_x\psi(x))
\label{scaleinvariance}
\eea
where $\phi(x)$ is real. Note that since the rescaling involves also
the matrix part of $D$, we must also rescale the masses of the
fermions. In this sense we differ form the usual usage of Weyl (or
conformal) invariance which is only valid for massless fields. In
our scheme Yukawa couplings are an integral part of the Dirac
operator which encodes all metric properties of the ``noncommutative
manifold'' described by the noncommutative matrix algebra. In the
absence of a dimensional scale, this is an exact symmetry of the
classical theory.

We now proceed to quantize the theory. It can be
proven~\cite{Fujikawabook} that if the classical theory is
invariant, the measure in the quantum path integral is not. We have
an anomaly: a classical theory is invariant against a symmetry
transformation, but the quantum theory, due to unavoidable
regularization, does not possess this symmetry anymore. If also the
quantum theory is required to be symmetric then the symmetry can be
restored by the addition of extra terms in the action,
alternatively one should have a fundamental length in the theory to explain
violation of Weyl invariance. A textbook introduction to anomalies can be found
in~\cite{Fujikawabook}. The notion of Weyl anomaly is attached to the
dilatation  of both coordinates, fields and mass-like parameters according to
their dimensionalities, Eq.~\eqn{scaleinvariance}. Evidently, in the absence of
UV divergences, there is no Weyl anomaly which therefore can be correlated to
rescaling of a cutoff in the theory. In the case when the dilatation is  not
constant, $\phi$ becomes a quantum field called the \emph{dilaton}. The dilaton
of this kind has been investigated in the context of the spectral action
in~\cite{ChamseddineConnesscale}.

We remark that there may be also an alternative realization of the
dilaton as a collective scalar mode of all fermions accumulated in a
{scale-noninvariant} dilaton action (in the spirit of~\cite{aano}).

We start from the partition function
\be
Z(D,\mu)=\int [\dd\psi] [\dd\bar\psi] e^{-S_\psi} =\det\left(\frac{D}{\mu}\right)\la{zinitial}
\ee
where we needed to introduce a normalization scale $\mu$ for
dimensional reasons, and the last equality is formal because the
expression is divergent and needs regularizing. The writing of the
fermionic action in this form (as a Pfaffian) is instrumental in the
solution of the fermion doubling problem in Connes approach to the
standard model~\cite{LMMS, G-BIS, AC2M2}.

In order to regularize the expression~\eqn{zinitial} we need to
introduce a \emph{cutoff} scale, which we call $\Lambda$. This is
the cutoff scale and it may have the physical meaning of an energy
in which the theory (seen as effective) has a phase transition, or
at any rate a point in which the symmetries of the theory are
fundamentally different (unification scale).

We then have two scales\footnote{In principle we would need also an
infrared regulator to render the spectrum of the Dirac operator
discrete. We will not discuss infrared issues here.}, and we will
keep them separated although in principle, at this stage, they could
be identified. We will see in the course of this work that they
cannot actually be identical, although have to be of the same order
of magnitude.

We will regularize the theory in the ultraviolet using a procedure
introduced by one of us, Bonora and Gamboa-Saravi
in~\cite{AndrianovBonoraGamboa, AndrianovBonora1, AndrianovBonora2}
but leaving room for the normalization scale $\mu$. Although this
procedure predates the spectral action, it is very much in the
spirit of spectral geometry, since it uses only the spectral data of
the Dirac operator. The energy cutoff is enforced by considering
only the first $N$ eigenvalues of $D$. Consider the projector
\be
P_N=\sum_{n=1}^N \ketbra{\lambda_n}{\lambda_n};\quad N=\max n \
\mbox{such that} \ \lambda_n\leq \Lambda \label{cuteigenvalues}
\ee
where $\lambda_n$ are the eigenvalues of $D$ arranged in increasing
order of their absolute value (repeated according to possible
multiplicities), $\ket{\lambda_n}$ a corresponding orthonormal
basis, and  the integer $N$ is a function of the cutoff. This means
that we are effectively using the $N^{\mathrm{th}}$ eigenvalue as
cutoff. Therefore this number and the corresponding spectral density
depends on coefficient functions of the Dirac operator, $N=N(D)$.

Instead of this sharp cutoff, which consider totally all eigenvalues
up to a certain energy, and ignore all the rest of the spectrum, it
is also possible to consider a smooth cutoff enforced by a smooth
function. Choosing a function $\chi$ which is smoothened version of
the characteristic function of the interval $[0,1]$ one can consider
the operator
\be
P_\chi=\chi\left(\frac D\Lambda\right)=
\sum_n \chi\left(\frac{\lambda_n}\Lambda\right)\  \ketbra{\lambda_n}{\lambda_n}  .
\ee
This operator is not a projector anymore, and it coincides with
$P_N$ for $\chi=\Theta$, where $\Theta$ is the Heaviside step
function.. The use of a smooth $\chi$ can be preferable in an
expansion, such as the heat kernel expansion we will perform later
in Sect.\ref{standard} for the spectral action. Nevertheless for the
scopes of the present paper a sharp cutoff is adequate.

In the framework of noncommutative geometry this is the most natural
cutoff procedure, although as we said it was introduced before the
introduction of the standard model in noncommutative geometry. It
makes no reference in principle to the underlying structure of
spacetime, and it is based purely on spectral data, thus is
perfectly adequate to Connes' programme. This form of regularization
could be also used for field theory which cannot be described on an
ordinary spacetime, as long as there is a Dirac operator, or
generically a wave operator, with a discrete spectrum.

We define the regularized partition function\footnote{Although $P_N$
commutes with $D$ we prefer to use a more symmetric notation.}
\bea
Z(D,\mu)&=&\prod_{n=1}^N\frac{\lambda_n}{\mu}
=\det\left(\one-P_N+P_N\frac{D}{\mu}P_N\right)\nonumber\\
&=& \det\left(\one-P_N+P_N\frac{D}{\Lambda}P_N\right)
\det\left(\one-P_N+\frac{\Lambda}{\mu}P_N\right)\nonumber\\
&=& Z_\Lambda(D,\Lambda)
\det\left(\one-P_N+\frac{\Lambda}{\mu}P_N\right). \label{3.7}
\eea
In this way we can define the fermionic action in an intrinsic way.

The regularized partition function $Z(D,\Lambda)$ has a well defined
meaning. Expressing $\psi$ and $\bar\psi$ as
\bea
\psi=\sum_{n=1}^\infty a_n\ket{\lambda_n};\qquad
\bar\psi=\sum_{n=1}^\infty b_n\ket{\lambda_n}
\eea
with $a_n$ and $b_n$ anticommuting (Grassman) quantities. Then
$Z(D,\Lambda)$ becomes (performing the integration over Grassman
variables for the last step)
\be
Z(D,\Lambda)=\int\prod_{n=1}^N \dfrac{\dd a_n \dd b_n}{\Lambda} \e^{-\sum_{n=1}^N b_n
\lambda_n
 a_n}=\det\left(D_N\right)
\ee
where we defined
\be
D_N=1-P_N+P_N\frac{D}{\Lambda}P_N .
\ee
In the basis in which $D/\Lambda$ is diagonal it corresponds to set
to $\Lambda$ all eigenvalues of $D$ larger than $\Lambda$. Note that
$D_N$ is dimensionless and depends on $\Lambda$ both explicitly and
intrinsically via the dependence of $N$ and $P_N$.

It is possible to give an explicit functional expression to the
projector in terms of the cutoff:
\be
P_N=\Theta\left(1-\frac{D^2}{\Lambda^2}\right)=\int\limits_{-\infty}^{\infty}
\dd\alpha\,\frac1{2\pi\ii(\alpha-
\ii\epsilon)} \e^{\ii\alpha\big(1-\frac{D^2}{\Lambda^2}\big)}
\ee
This integral is well defined for a compactified space volume.
Actually $N$ depends also on the infrared cutoff, and the number of
dimensions.

\section{Bosonic Action from Weyl Anomaly}

In this section we will see how the Weyl anomaly induces the bosonic
part of the action. The induced action is the Chamseddine-Connes
spectral action.

The action $S_\psi$ is invariant under~\eqn{scaleinvariance}, but
the partition function \eqn{zinitial} is not. The reason for this is
the fact that the regularization procedure is not Weyl invariant.
In~\cite{anlizzi} it was shown that the anomaly can in principle be
absorbed by a change of the measure, which is equivalent to the
addition of another term to the action. This term can compensate the
change in the measure due to the regularization, but being in an
exponential form, can also be seen as another addition to the
action, so that the final partition function is invariant. This
calculation has been originally performed in~\cite{AANN} in the QCD
context, and applied to gravity in~\cite{NovozhilovVassilevich}. In
the scenario we are favoring in this paper however Weyl symmetry is
not an exact symmetry of the theory, and the bosonic part of the
action is induced by the renormalization group flow.

In the following we will mostly consider the case of $\phi$ constant
(i.e.\ not depending on $x$). This simplifies things because we do
not have to worry about the kinetic terms of the field, and renders
the functional integrals simple integrals.

In order to make contact with the spectral action (to be discussed
next) let us notice that $N$ is just the number of eigenvalues
smaller that $\Lambda$, and thereby
\be
\Tr\chi\left(\frac{D^2}{\Lambda^2}\right)=
\Tr\Theta\left(1-\frac{D^2}{\Lambda^2}\right)=\Tr P_N= N(\Lambda,\ D). \label{spact}
\ee
where $\chi$ is a generic cutoff function, which in our case is a
sharp cutoff at energy $\Lambda$,
\be
\chi(x)=\left\{\begin{array}{cc}0~ & x<0\\ 1~ & x\in[0,1]\\ 0~ & x>1
\end{array}\right. \label{sharpcutoff}
\ee
consequence of the sharp cutoff on the eigenvalues used
in~\eqn{cuteigenvalues}. For smoother cutoffs of the eigenvalues
this would reflect in different forms of $\chi$. We will see in the
next section that at one loop level (the only doable approximation)
the actual form of the cutoff is not crucial. The latter form
or~\eqn{spact} is valid provided that we take into account the
functional dependence $N=N(\Lambda,\ D)$. It is worth recalling
again that the integer $N$ depends on the cutoff $\Lambda$, on the
Dirac operator $D$ and also on the function $\chi$ which we have
chosen to be a sharp cutoff.

If we want to obtain a partition function invariant on $\phi$ we can
integrate it out, i.e.
\be
Z_{\mathrm{inv}}(D,\mu)=\int\dd\phi Z(\e^{-\frac12\phi}
D\e^{-\frac12\phi}, \mu)\equiv \int\dd\phi Z( D_\phi, \mu);\qquad
D_\phi\equiv \e^{-\frac12\phi} D\e^{-\frac12\phi} . \label{caseA}
\ee
This was the procedure followed in~\cite{anlizzi}. Notice however
that in principle we could have equally well defined
\be
\hat Z_{\mathrm{inv}}(D, \mu)= \left(\int\dd\phi \dfrac{1}{ Z (
D_\phi, \mu)}\right)^{-1} . \label{caseA1}
\ee
If we consider non Weyl invariant partition function we can split
it in the product of a term invariant for Weyl transformations, and
another not invariant, which will depend on the field $\phi$.
\be
Z(D,\mu)=\hat{Z}_{\mathrm{inv}}(D,\mu)Z_{\mathrm{not}}(D,\mu)
\label{splitting}
\ee
The terms in $Z_{\mathrm{not}}$ are due to the Weyl anomaly and we
can calculate them. Using
\be
D_\phi=\e^{-\frac12\phi}D\e^{-\frac12\phi}
\ee
consider the identity
\be
Z(D)=\left(\int[d\phi]\frac1{Z(D_\phi)}\right)^{-1}\, \int [d\phi]
\frac{Z(D)}{Z(D_\phi)}
\ee
Since the first term is invariant by construction, the second is the
not invariant one:
\be
Z_{\mathrm{not}}(D)=\int[d\phi]e^{-S_{\mathrm{not}}}=\int [d\phi]
\frac{Z(D)}{Z(D_\phi)}
\ee

To obtain the Weyl invariant partition function we need to multiply the
regularized one by a compensating term, which we express in exponential form,
as an addition to the action which we call the anomalous action.
\be
{Z_{\mathrm{inv}}}(D,\mu)=Z(D,\mu)\cdot
Z_{\mathrm{anom}}(D,\mu);\quad Z_{\mathrm{anom}}(D,\mu) =
\int\dd\phi\, \e^{-S_{\mathrm{anom}}} \la{zinvprod}
\ee
where the effective action will be depending on $N$, and hence the
cutoff $\Lambda$, and on $\phi$. Then
\be
S_{\mathrm{anom}}= \log\left( \dfrac{Z (D, \mu)}{Z( D_\phi, \mu)}\right) \label{sanom1}
\ee

Notice that the splitting in~\eqn{splitting} is of course not
unique, but is motivated by the following. We know
from~\cite{anlizzi} that if we add to the classical action the term
$S_{\mathrm{anom}}$ we will restore the Weyl invariance of the
partition function $Z$. Thereby it is essential to have
$S_{\mathrm{not}} = -S_{\mathrm{anom}}$. We shall see below
(Eq.~\eqn{sanomcoll}), that the choice~\eqn{caseA1} provides such an
equality.

Let us define
\be
Z_t=Z(D_{t\phi}, \mu)
\ee
therefore $Z_0=Z(D, \mu)$ and
\be
\dfrac{Z_{\mathrm{inv}}(D, \mu)}{Z(D, \mu)}=\int\dd\phi \frac{Z_1}{Z_0}
\ee
and hence
\be
S_{\mathrm{anom}}=-\int_0^1\dd t \del_t \log Z_t =-\int_0^1\dd t
\frac{\del_t Z_t}{Z_t}
\ee
We have the following relation that can easily proven
\bea
\del_t Z_t=\del_t\det \left(\frac{D_{t\phi}}{\mu}\right)_N
&=&\phi Z_t  \left(-1 +\Lambda^2 \log\frac{\Lambda^2}{\mu^2} \partial_{\Lambda^2}\right) \tr P_N ,
\eea
and therefore, for $\phi$ not dependent on $x$,
\bea
S_{\mathrm{anom}}&=&\int_0^\phi\dd t'\,
\left(1 -\Lambda^2 \log\frac{\Lambda^2}{\mu^2}
\partial_{\Lambda^2}\right)\Tr\Theta\left(1-\frac{D_{t'}^2}{\Lambda^2}\right)\nonumber\\
&=& \int_0^\phi\dd t'\,  \left(1 -\Lambda^2
\log\frac{\Lambda^2}{\mu^2} \partial_{\Lambda^2}\right) N(\Lambda,\
D_{t'}).\ \label{Sanomal}
\eea

The presence of the bosonic action given by the trace of the
regularized Dirac operator is a consequence of the renormalization
flow of the partition function. Under the change
\be
\mu\to\gamma\mu
\ee
with $\gamma$ real. From~\eqn{3.7} the partition function changes as
follows
\be
Z(D,\mu)\to Z(D,\mu)e^{-(\log\gamma)\tr P_N}
\ee
and
\be
\tr P_N=N=\tr\chi\left(\frac D\Lambda\right) \label{inducedaction}
\ee
as always for the choice of $\chi$ the characteristic function on
the interval, a consequence of our sharp cutoff on the eigenvalues.

The expression~\eqn{inducedaction} is nothing but the spectral
action which we will discuss in the next section. We see therefore
that the renormalization group flow of the fermionic action induces
the bosonic spectral action. The anomalous part of the
action~\eqn{Sanomal} is a modification of the
action~\eqn{inducedaction}.

\section{The Spectral Action Principle \label{Connesspectral}}

In this section we give a briefest introduction to the relevant
aspects of the spectral action principle. The reader conversant with
the topic may skip this section. More thorough introduction can be
found in~\cite{Schucker,AC2M2,CCintro}.

\subsection{Fields, Hilbert Spaces, Dirac Operators and the (Non)commutative
Geometry of Spacetime}

The main idea of the whole programme of Connes' noncommutative
geometry~\cite{Connesbook} is to describe ordinary mathematics, and
physics, in term of the spectral properties of operators. This
programme has its roots in quantum mechanics and aims at the
description of generalized spaces. The main ingredients are an
algebra represented on a Hilbert space, and the generalized Dirac
operator which describes all metric aspects of the theory, and as we
have seen the behavior of the fundamental matter fields, represented
by vectors of the Hilbert space. The fluctuations of the Dirac
operator instead contain all boson fields, including the mediators of the
forces (intermediate vector bosons), and the Higgs field.

We have introduced a (Euclidean) spacetime. And therefore implicitly
the algebra $\cal A$ of complex valued continuous functions of this
space time. There is in fact a one-to one correspondence between
(topological Hausdorff) spaces and commutative $C^*$-algebras, i.e.\
associative normed algebras with an involution and a norm satisfying
certain properties. This is the content of the Gelfand-Naimark
theorem~\cite{FellDoran, Ticos}, which describes the topology of
space in terms of the algebras. In physicists terms we may say the
the properties of a space are encoded in the continuous fields
defined on them. This concept, and its generalization to
noncommutative algebras is one of the starting points of Connes'
noncommutative geometry programme~\cite{Connesbook}. The programme
aims at the transcription of the usual concepts of differential
geometry in algebraic terms and a key role of this programme is
played by a \emph{spectral triple}, which is composed by an algebra
acting as operators on a Hilbert space and a (generalized) Dirac
operator. In our case we have these ingredients, but we have to
consider instead of the the algebra of continuous complex valued
function, matrix valued functions. The underlying space in this case
is still the ordinary spacetime, technically the algebra is ``Morita
equivalent'' to the commutative algebra, but the formalism is built
in a general way so to be easily generalizable to the truly
noncommutative case, when the underlying space may not be an
ordinary geometry.

The spectral triple contains the information on the geometry of spacetime. The
algebra as we said is dual to the topology, and the Dirac operator enables the
translation of the metric and differential structure of spaces in an algebraic
form. There is no room in these proceedings to describe this programme, and we
refer to the literature for details~\cite{Connesbook, Landibook, Ticos,
Madore}.

Within this general programme a key role is played by the approach
to the standard model. This  is the attempt to understand which kind
of (noncommutative) geometry gives rise to the standard model of
elementary particles coupled with gravity. The roots of this
approach is to have the Higgs appear naturally as the ``vector''
boson of the internal noncommutative degrees of
freedom~\cite{Madoreearly, D-VKM, ConnesLott}. The most complete
formulation of this approach is given by the \emph{spectral action},
which in its most recent form is presented in~\cite{AC2M2}.

\subsection{The Spectral Action and the Standard Model coupled to Gravity \label{standard}}

The integrand in \eqref{Sanomal} is basically the Chamseddine-Connes Spectral
Action introduced in~\cite{SpectralAction} together with the
fermionic action~\eqn{fermionicaction}. More precisely the bosonic
part of the spectral action is
\be
S_B = \Tr \chi\left(\frac{D^2}{\Lambda^2}\right)
\ee
The bosonic spectral action so introduced is always finite by its
nature, it is purely spectral and it depends on the cutoff
$\Lambda$. For the choice of $\chi$ as sharp cutoff we have that
the trace counts exactly the eigenvalues smaller than $N$, and
therefore
\be
S_B=N(D,\Lambda)
\ee

In the original work of Chamseddine and Connes the bosonic and
fermion parts of the action were treated differently. The fermionic
action on the contrary is divergent, and will require
renormalization.
We have seen as the cancelation of the anomaly brings the two
actions on the same footing, albeit with a modification of the
bosonic part. We notice that already in~\cite{Sitarz} the two
actions were proposed to ``unify'' in the bosonic action with the
addition of the projection on the fermionic field to the covariant
Dirac operator. This reproduces the full spectral action with some
additional non linear terms for the fermions, which could have to do
with fermionic masses. Recently Barret~\cite{Barrett} has argued
that the bosonic spectral action can inferred from the fermionic
action via the state sum model. His work has some points of contact
with ours.

To obtain the standard model take as algebra the product of the
algebra of functions on spacetime times a {finite dimensional}
matrix algebra
\be \mathcal A =C({\mathbb
R}^4)\otimes{\mathcal A}_F
\ee
Likewise the Hilbert space is the product of fermions times a finite
dimensional space which contains all matter degrees of freedom, and
also the Dirac operator contains a continuous part and a discrete
one
\be
\mathcal H ={\mathrm{Sp}({\mathbb R}^4)\otimes{\mathcal H}_F}
\ee
and the Dirac operator
\be
D_0=\gamma^\mu\del_\mu\otimes\mathbb I + \gamma\otimes D_F
\ee
In its most recent form  due to Chamseddine, Connes and
Marcolli~\cite{AC2M2} a crucial role is played by the mathematical
requirements that the noncommutative algebra satisfies the
requirements to be a manifold. Then the internal algebra, is almost
uniquely derived to be
\be
{\mathcal A}_F=\complex\oplus{\mathbb H}\oplus M_3(\complex)
\ee
Then the  bosonic spectral action can be evaluated at one loop using
standard heath kernel techniques~\cite{Vassilevich:2003xt} and the
final result gives the full action of the standard model coupled
with gravity. We restrain from writing it since it takes more than
one page in the original paper~\cite{AC2M2}. In the process however
one does not need to input the mass of the Higgs, which comes out as
a prediction. Its value comes out to be $\sim 170 \mathrm{GeV}$. A
small value experimentally disfavored. It must be said however that
the present form of the model needs unification of the three
coupling constant at a single energy point (given by $\Lambda$). The
model also contains nonstandard gravitational terms (quadratic in
the curvature), which are currently being investigated for their
cosmological consequences~\cite{NelsonSakellariadu,
MarcolliPierpaoli}.

Technically the canonical bosonic spectral action is a sum of residues, and
can be expanded in a power series in terms of $\Lambda^{-1}$ as
\be
S_B(\Lambda)=\sum_n f_n\, a_n(D^2/\Lambda^2)
\ee
where the $f_n$ are the momenta of $\chi$
\begin{eqnarray}
f_0&=&\int_0^\infty \dd x\, x  \chi(x)\nonumber\\
f_2&=&\int_0^\infty \dd x\,   \chi(x)\nonumber\\
f_{2n+4}&=&(-1)^n \del^n_x \chi(x)\bigg|_{x=0} \ \ n\geq 0
\label{fcoeff}
\end{eqnarray}
the $a_n$ are the Seeley-de Witt coefficients which vanish for $n$
odd. For $D^2$ of the form
\be
D^2=-(g^{\mu\nu}\del_\mu\del_\nu\one+\alpha^\mu\del_\mu+\beta)
\ee
defining
\begin{eqnarray}
\omega_\mu&=&\frac12 g_{\mu\nu}\left(\alpha^\nu+g^{\sigma\rho} \Gamma^\nu_{\sigma\rho}\one\right)\nonumber\\
\Omega_{\mu\nu}&=&\del_\mu\omega_\nu-\del_\nu\omega_\mu+[\omega_\mu,\omega_\nu]\nonumber\\
{\mathcal E}&=&\beta-g^{\mu\nu}\left(\del_\mu\omega_\nu+\omega_\mu\omega_\nu-\Gamma^\rho_{\mu\nu}\omega_\rho\right)
\end{eqnarray}
then
\begin{eqnarray}
a_0&=&\frac{\Lambda^4}{16\pi^2}\int\dd x^4 \sqrt{g}
\tr\one_F\nonumber\\
a_2&=&\frac{\Lambda^2}{16\pi^2}\int\dd x^4 \sqrt{g}
\tr\left(-\frac R6+{\mathcal E}\right)\nonumber\\
a_4&=&\frac{1}{16\pi^2}\frac{1}{360}\int\dd x^4 \sqrt{g}
\tr(-12\nabla^\mu\nabla_\mu R +5R^2-2R_{\mu\nu}R^{\mu\nu}\nonumber\\
&&+2R_{\mu\nu\sigma\rho}R^{\mu\nu\sigma\rho}-60R{\mathcal E}+180{\mathcal E}^2+60\nabla^\mu\nabla_\mu
{\mathcal E}+30\Omega_{\mu\nu}\Omega^{\mu\nu}) \label{spectralcoeff}
\end{eqnarray}
$\tr$ is the trace over the inner indices of the finite algebra
$\mathcal A_F$ and in $\Omega$ and $\mathcal E$ are contained the gauge
degrees of freedom including the gauge stress energy tensors and the
Higgs, which is given by the inner fluctuations of $D$.

In our case
for $\phi$ constant, after performing the integration we find
\bea
S_{\mathrm{anom}}&=& \left(1 -\Lambda^2 \log\frac{\Lambda^2}{\mu^2} \partial_{\Lambda^2}\right) \int_0^\phi \dd t'  S_B (\Lambda e^{t'})
\nonumber\\&=& \left(1 -\Lambda^2 \log\frac{\Lambda^2}{\mu^2} \partial_{\Lambda^2}\right) \int_0^\phi \dd t' \sum_n \e^{(4-n)t'} a_n f_n\nonumber\\&=&
\frac{1}{8} (e^{4\phi}-1) a_0\left(1 - 2~ \log\frac{\Lambda^2}{\mu^2} \right) + \frac{1}{2} (e^{2\phi}-1) a_2 \left(1 -  \log\frac{\Lambda^2}{\mu^2} \right)+ \phi
a_4 .\label{Sanomaaction}
\eea
There are just  some numerical corrections to the first two
Seeley-de Witt coefficients due to the integration in $t' = t\phi$
and a choice of normalization scale $\mu$. In the case of a non
sharp cutoff some numerical coefficients would change according
to~\eqn{fcoeff}, and of course the series would not terminate at
$a_4$. The corrections are however small, and the remaining terms
are subdominant. Therefore the presence of a different cutoff would
not alter the qualitative aspects of what follows.

The sign with which this action appears in he partition function is
of course crucial. We will see in the next section of the
interpretation of $\phi$ as emerging from bosonization choices a
sign. And later on in Sect.~\ref{se:dilatoneffpot} we will see that
the sign chosen in this case gives a qualitative realistic effective
Higgs-dilaton potential.


\section{Dilaton bosonization}

In this section we will discuss the role of the dilaton considering
it as arising from a bosonization process of high energy degrees of
freedom. We are interested in the effective potential, therefore we
will make the brutal assumption of considering only a
\emph{constant} (i.e. not dependent on $x$) dilaton $\phi$ and Higgs
field $H$. In this case $H$ is the only surviving term in the
off-diagonal entries of~\eqn{dirac22}. In fact here by ``Higgs
field'' we mean generically all degrees of freedom which connect left
and right chiralities. The analysis carried is therefore quite solid
and independent on the details of the model.

The action after bosonization can be represented as,
\be
Z(D, \mu)=\hat Z_{\mathrm{inv}}(D, \mu) \int\dd\phi\,
\e^{-S_{\mathrm{\mathrm{coll}}}} \label{caseB}
\ee
then
\be
S_{\mathrm{\mathrm{coll}}}= \log\left(\frac{Z( D_\phi, \mu)}{Z(D,
\mu)}\right) = - S_{\mathrm{anom}},\la{sanomcoll}
\ee
which is to be confronted with~\eqref{sanom1}.

The Higgs mechanism of spontaneous symmetry breaking is not
compatible with the Weyl conformal invariance. Indeed, let us
consider the dependence of the invariant partition function
$Z_{\mathrm{inv}}$, given by (\ref{caseA}) or \eqref{caseA1}, on the
Higgs field $H$.
\begin{equation}
Z_{\mathrm{inv}}  =  e^{-W_{\mathrm{inv}}(H,g_{\mu\nu,...})}\label{inv1},
\end{equation}
where
\begin{equation}
W_{\mathrm{inv}} = \int d^4 x \sqrt{|g|} (\lambda H^4  + \mbox{\rm terms with derivatives}) \label{inv2}.
\end{equation}
We omit in the righthand side of~(\ref{inv2}) terms with derivatives
of the Higgs fields and powers of the Riemann curvature tensor
because in this work we are concerned only by properties of the
effective potential for Higgs and dilaton fields. The form of
$W_{\mathrm{inv}}$ could be guessed by dimensional analysis, but we
show in the appendix how it emerges (with the correct sign).
In~\eqref{inv2} there are no terms generating spontaneous symmetry
breaking to supply Higgs fields with a mass and accordingly we
assume, that the Higgs field mass formation is related to the Weyl
noninvariant part of the partition function. The latter one is
determined by conformal anomaly term~\eqref{Sanomaaction}. Let us
investigate how the composite dilaton field is related to the
primary fields of the theory under consideration:
$\psi,\bar{\psi},H$. For a fixed configuration of the Higgs field
$H$, the dilaton field $\varphi$ appears as a result of bosonization
of the fermions $\psi,\bar{\psi}$.

The bosonization is defined by identifying
\begin{equation}
Z_{\mathrm{fermion}}(j) = Z_{\mathrm{boson}}(j) \label{bosonisation},
\end{equation}
where
\begin{eqnarray}
&&Z_{\mathrm{fermion}}(j) = \int D \psi D\bar{\psi} e^{-W(\psi,\bar{\psi},j)},\nonumber\\
 &&Z_{\mathrm{boson}} (j)\simeq \int D \varphi e^{-S_{\mathrm{\mathrm{coll}}}(\varphi,j) - W_{\mathrm{inv}}(j)}=
e^{-W_{\mathrm{\mathrm{coll}}}(j) - W_{\mathrm{inv}}(j)},  \label{Zfb}
\end{eqnarray}
Herein $j$ denotes a set of sources. In our bosonization scheme the Higgs field $H$,
is  treated as a source for a scalar combination $\psi\bar{\psi}$ and therefore it is fixed in the process of dilaton bosonization (included in to $j$).  In the definition of the bosonic partition function  "$\simeq$" signifies that  we neglect the bosonic fields with the spin more than 0 and retain only one scalar (dilaton) degree of freedom. We have already seen that  the Higgs mechanism is presumably  related to the violation of the Weyl symmetry and the latter is given by conformal anomaly term \eqref{Sanomaaction}, which exploits only one zero-spin field besides the Higgs field itself. Thus we conclude that our simplification is reasonable.  In order to investigate full dynamics, one should deal with the total partition function
\begin{equation}
Z_{\mathrm{total}} = \int D H \left(Z_{\mathrm{fermion\ or\ boson}}(H,\tilde{j})\right),
\end{equation}
where $\tilde{j}$ is a set of sources for all quantum fields,
excluding $H$.

Varying both the left and the right hand sides of
(\ref{bosonisation}) over $H$, one derives the equation, that
relates fermion condensate $\langle  \psi \bar{\psi} \rangle$ with
the average values (over bosonic vacuum) of the combination of the
bosonic fields $H$ and $\phi$,
\begin{equation}
\langle  \psi \bar{\psi} \rangle \propto \frac{\delta \ln Z_{boson}(H)}{\delta H} =
-\left\langle \frac{\delta (S_{\mathrm{\mathrm{coll}}}(H,\varphi)+W_{\mathrm{inv}}(H))}{\delta H}
\right\rangle .
\end{equation}
This relation allows to unravel the bosonic content of fermion
bilinear operator in different phases: symmetric with $\langle
H\rangle = 0$ and symmetry breaking one with $\langle H\rangle \not=
0$.

Thus the two different choices of dilaton field correspond to two
different interpretations of the $\phi$ degree of freedom. The
different choices are described in the definitions~\eqn{caseA}
and~\eqn{caseB}. From these descend the definition of the
alternative $Z_{\mathrm{inv}}$ or $\hat Z_{\mathrm{inv}}$. The
former choice~\eqn{caseA} is the natural one if one has a
noninvariant partition function and wants to define an invariant one
by including an extra fundamental degree of freedom. The latter
choice is instead the natural one in the case in which one starts
from a non invariant theory in which the dilaton is a composite
object whose condensates restores a global symmetry. This bosonic
degree can be some fermionic bilinear. In the following we will give
some arguments in favor of this second choice, based on the
interplay with the Higgs field.
\section{The Dilaton and the effective potential \label{se:dilatoneffpot}}
The full analysis of the model coupled with a dynamical dilaton is
under way and will be published elsewhere. Nevertheless it is
already possible to say something on the interplay between the
dilaton and the Higgs, and in particular the effective potential.
This can be used to characterize cosmic evolution right after
inflation starts. In particular, it may open the ways to describe
the transition from the radiation phase with massless particles to
the EW symmetry breaking phase with spontaneous mass generation due
to condensation of Higgs fields.

\subsection{Mass generation from Higgs-dilaton potential during
cosmic evolution} We will consider in the following only the
potential terms relative to the complex Higgs doublet $H$ and the
dilaton $\phi$.


Because of Weyl invariance, within our approximations, the only
allowed dependence on the Higgs field H of $\hat Z_{\mathrm{inv}}$
in \eqref{caseB} is given by(see Eqs.~\eqref{inv1} and \eqref{inv2})
\be
\hat Z_{\mathrm{inv}} = e^{-\hat
W_{\mathrm{inv}}(H,g_{\mu\nu},...)},
\ee
where
\begin{equation}
\hat W_{\mathrm{inv}} = \int d^4 x \sqrt{|g|} (-C\phi_0 H^4
  + \mbox{\rm terms with derivatives})
 \label{WINV},
\end{equation}
with some (not yet defined) constant $\phi_0$ and  $C$ is fixed
positive constant, which we define below in Eq.~\eqref{C}. Later we
will see, that only the choice $\phi_0 < 0$ supports the spontaneous
EW symmetry breaking. Let us define the effective Higgs-dilaton
potential $V$ by the equality
\be
Z(D) = \int D\phi ~e^{-\hat W_{\mathrm{inv}}(H) -
S_{\mathrm{coll}}(H,\phi)}\equiv \int D\phi ~e^{-\int d^4 x
\sqrt{|g|}V(H,\phi)}.\la{Zscen2}
\ee
We can derive the form of effective Higgs-dilaton potential. To
focus on this goal we reduce the joint effective Higgs-dilaton (HD)
potential including only the real scalar component $H$ of the
(complex) Higgs doublet $ (H_1, H_2) \to (0, H) $ subject to
condensation. From the expression \eqref{Sanomaaction} for
$S_{\mathrm{anom}}$ one obtains the following formula for the  the
effective Higgs-dilaton potential $V$: \ba
V &=& V_{\mathrm{coll}} + \hat W_{\mathrm{inv}},\label{01}, \\
V_{\mathrm{coll}} &=& \tilde A \left(e^{4\phi}-1\right) + \tilde B
H^2\left(e^{2\phi} - 1\right) - CH^4\phi. \la{vcoll} \ea The
explicit form of the coefficient is given in the appendix.

The quadratic term of the Higgs potential comes from the $a_2$ term
of \eqn{Sanomaaction}, while the quartic one comes from the $a_4$
one and from $\hat W_{\mathrm{inv}}$. Evidently the constant
$\phi_0$ can be eliminated by shifting the field $\phi \to
\phi-\phi_0$ and rescaling the constants $\tilde A,\tilde B$.
 After performing renormalization the
general form of the HD potential can be presented as,
\be
V = Ae^{4\phi} + BH^2e^{2\phi} - C\phi H^4 + EH^2 + V_0,\la{1}
\ee
where
\ba
A &=& \frac{45}{8\pi^2}\frac{\Lambda^4}{8}\left(2\log\frac{\Lambda^2}{\mu^2}-1\right)e^{-4\phi_0},\la{Aa}\\
B &=& \frac{3 y^2}{2\pi^2}\frac{\Lambda^2}{2}\left(1 - \log\frac{\Lambda^2}{\mu^2}\right)e^{-2\phi_0},\la{Ba} \\
C &=& \frac{3 z^2}{4\pi^2},\la{Ca}\\
E &=&-\frac{3 y^2}{2\pi^2}\frac{\Lambda^2}{2}\left(1 - \log\frac{\Lambda^2}{\mu^2}\right),\la{E}\\
V_0 &=&
-\frac{45}{8\pi^2}\frac{\Lambda^4}{8}\left(2\log\frac{\Lambda^2}{\mu^2}-1\right).\la{V0}
\ea In the formulas \eqref{Ba},\eqref{Ca},\eqref{E} the constants
$y$ and $z$ depend on mixing an Yukawa couplings. Their exact
definition is in \cite[Formula 3.17]{SpectralAction}. In~\eqref{1}
depending on the normalization scale $\mu$ of the fermion effective
action, compared with the cutoff $\Lambda$, one can get in principle
any sign of the coefficients $A(\Lambda,\mu), B(\Lambda,\mu)\gtrless
0 $. Thus in general both signs and modules of these constants $A,B$
are possible..

Here we are interested in the evolution of fields $\phi$ and $H$ and
correspondingly neglect the additional cosmological constant $V_0$.
We would like to apply the HD potential in the framework of the
description of cosmic evolution. This evolution will depend
principally on the signs of the constants, and on relations among
their modules. We therefore search for which combinations of signs
can provide the evolution from a symmetric phase to the EW symmetry
broken phase, with the generation of fermion mass due to the Higgs
fields. Thus one has to inquire whether the HD potential has local
minima, and what are the restrictions on the coefficients which
provide the existence of such minima. Accordingly we are going to
investigate all possible critical points\footnote{Here by critical
point we mean a stationary one.} of this potential depending on the
values of its coefficients. The potential~\eqref{1} has three
arbitrary parameters $A,B,E$, but it must be sign $(B)$ =
-sign$(E)$. The parameter $C$ is fixed and given by \eqref{Ca}.
Nevertheless in the analysis of extremal properties of $V$ performed
below we shall consider arbitrary $C, B$ and $E$. We will see that,
in order to have symmetry breaking, indeed the constant $C$ must be
positive, and $E$ and $B$ must have opposite signs. This is a
confronting result.

Without loss of generality one can impose $C>0$. For the opposite
sign of $C$ the set of critical points can be found by reflection
$V\to - V$. One can see, that $V$ has no any critical points at
$H=0$. Let us perform the coordinate transformation to the variable
$\eta$, \be H^2 = \eta e^{2\phi}\label{13} \ee

Such a transformation is non-degenerate at $H \neq 0$ and, since $V$
is symmetric for $H\to-H$, preserves all the information about
extremal properties of our potential.

In the new variables the potential takes the form,
\be
V = e^{4\phi}\left(A + B\eta - C \phi \eta^2\right) + E e^{2\phi}
\eta . \label{4}
\ee
Critical point coordinates obey the following equations, \ba
2A + B \eta  - \frac{C}{2} \eta^2   &=& 0 \label{5}\\
\left(\frac{2C\eta}{E}\right)\phi  - \frac{B}{E} &=&  e^{-2\phi}
\label{6} \ea with the additional requirement $\eta > 0$ .

From the equation (\ref{5}) we immediately find,
\be
\eta_{1,2} = {\frac {4A}{-B\pm\sqrt {{B}^{2}+4\,AC}}} .\label{8}
\ee
It is known (for a quick introduction see e.g.~\cite{wiki}), that
the equation of a type $a x + b = p^{c x + d}~~a,c\neq 0$, can be
exactly solved in terms of the Lambert $W(z)$
function~\cite{lambert}. By definition, it is a solution of the
equation,
\be
z = W(z)e^{W(z)} \label{lam}
\ee
The function $W e^W$ is not injective and  $W$ is multivalued
(except at 0). If we look for real-valued $W$ then the relation
\eqref{lam} is defined only for $x \geq 1/e$, and is double-valued
on $(-1/e, 0)$.

Let us introduce the notation $W_{0}(x)$ for the upper branch. It is
defined at $-1/e \leq x < \infty$ and it is monotonously increasing
from -1 to $+\infty$. The lower branch is usually denoted
$W_{-1}(x)$. It is defined only on $-1/e\leq x< 0$ and it is
monotonously decreasing from -1 to $-\infty$.

In these terms the general solution of \eqref{6} is given by,
\be
\phi = \frac12 W \left( \frac{E e^{-\frac{B}{\eta C}}}{\eta C}
\right) +{\frac {B}{2\eta\,C}}
 \label{10}
\ee
Since we have two values of $\eta$ and the real $W$ is
double-valued, then the maximal number of critical points is four.
However $\eta$ must be positive and real, and $\phi$ must be real.
From these requirements one obtains the restrictions on the
coefficients, which provide an existence of each critical point.

We shall denote our critical points as $(m,n)$. Here the first index
$m$ marks the sign $\pm$ and corresponds to the type of a chosen
$\eta$ from~(\ref{8}). The index $n$ ranges over $-1,0$ and
corresponds to the chosen branch of the $W$ function. We specify a
type of each critical point with the help of the Hessian matrix
eigenvalues and find the following results for the acceptable
composition of coefficient signs.

We seek for combinations of signs of the coefficients $A,B,C,E$
which  provide a \emph{minimum} triggering the spontaneous EW
symmetry breaking at a final stage of cosmic evolution. There are 11
combinations of signs  which are forbidden as they don't  provide
the existence of a local minimum.
\begin{table}
\begin{center}\begin{tabular}{|c|c|c|c|}\hline
sign (A)&sign(B)& sign(C)&sign(E)\\\hline $\pm$ & $\pm$& +&+\\\hline
-&-&+&-\\\hline -&$\pm$&-&$\pm$\\\hline +&+&-&$\pm$\\\hline
\end{tabular}
\end{center}
\caption{\sl Choice of signs which \emph{do not} give a local
minimum to the potential. \label{Tab1}}
\end{table}
The only five combinations of signs which give the required minimum
are shown in Table~\ref{Tab2}.
\begin{table}[htp]
\begin{center}\begin{tabular}{|c|c|c|c|}\hline
sign (A)&sign(B)& sign(C)&sign(E)\\\hline + & $+$& +&-\\\hline
+&-&+&-\\\hline -&$+$&+&-\\\hline +&-&-&$+$\\\hline
+&-&-&$-$\\\hline
\end{tabular}
\end{center}
\caption{\sl Choice of signs which \emph{do} give a local minimum to
the potential. \label{Tab2}}
\end{table}

\subsection{Transition from symmetric phase to Electroweak symmetry breaking
phase  and choice of signs}
We now examine the possibility of scenario where, at the first stage
of the Universe evolution, one deals with massless fermions with the
vanishing vacuum expectation value of the Higgs field $\langle
H_{in} \rangle = 0$ (symmetric phase). We consider an initial point
$(\phi_{in}, H_{in} = 0)$ acceptable for starting evolution if the
function $V|_{\phi = \phi_{in}}(H)$ has a local minimum at $H = 0$,
and if we can roll down from the initial point to a final one which
is a local minimum corresponding to the Higgs phase. We have listed
in table~\ref{Tab2} the five combinations of signs of the parameters
$A$, $B$, $C$, $E$ which provide the existence of the local minimum
. Nevertheless not all of these combinations support the above
transition scenario. Indeed one can prove that this scenario can be
realized only for positive $A,B,C$ and negative $E$. For this case
the solution for minimum belongs to the class $(+,-1)$ and the
minimum (final-stage) coordinates  are given by, \ba
\eta_{fin} &=& {\frac {4A}{-B+\sqrt {{B}^{2}+4\,AC}}} > 0\label{20},\\
\phi_{fin} &=& \frac12 W_{-1} \left( \frac{E e^{-\frac{B}{\eta_{fin}
C}}}{\eta_{fin} C} \right) +{\frac {B}{2\eta_{fin}\,C}} .\label{21}
\ea The requirement for $\phi$ to be real leads to,
\be
{E_{min} < E < 0,~~~~~E_{min}\equiv -
C\eta_{fin}\exp\left\{ -1 + \frac{B}{\eta_{fin} C} \right\}}
\label{c6}
\ee
The additional bounds exist on the coefficients,
\be
B e^{2\phi_{in}} + E > 0,
\ee
to guarantee that the initial point is in the symmetric phase.
\begin{figure}[htb]
\vskip -0.5cm
\includegraphics
[scale=.2]{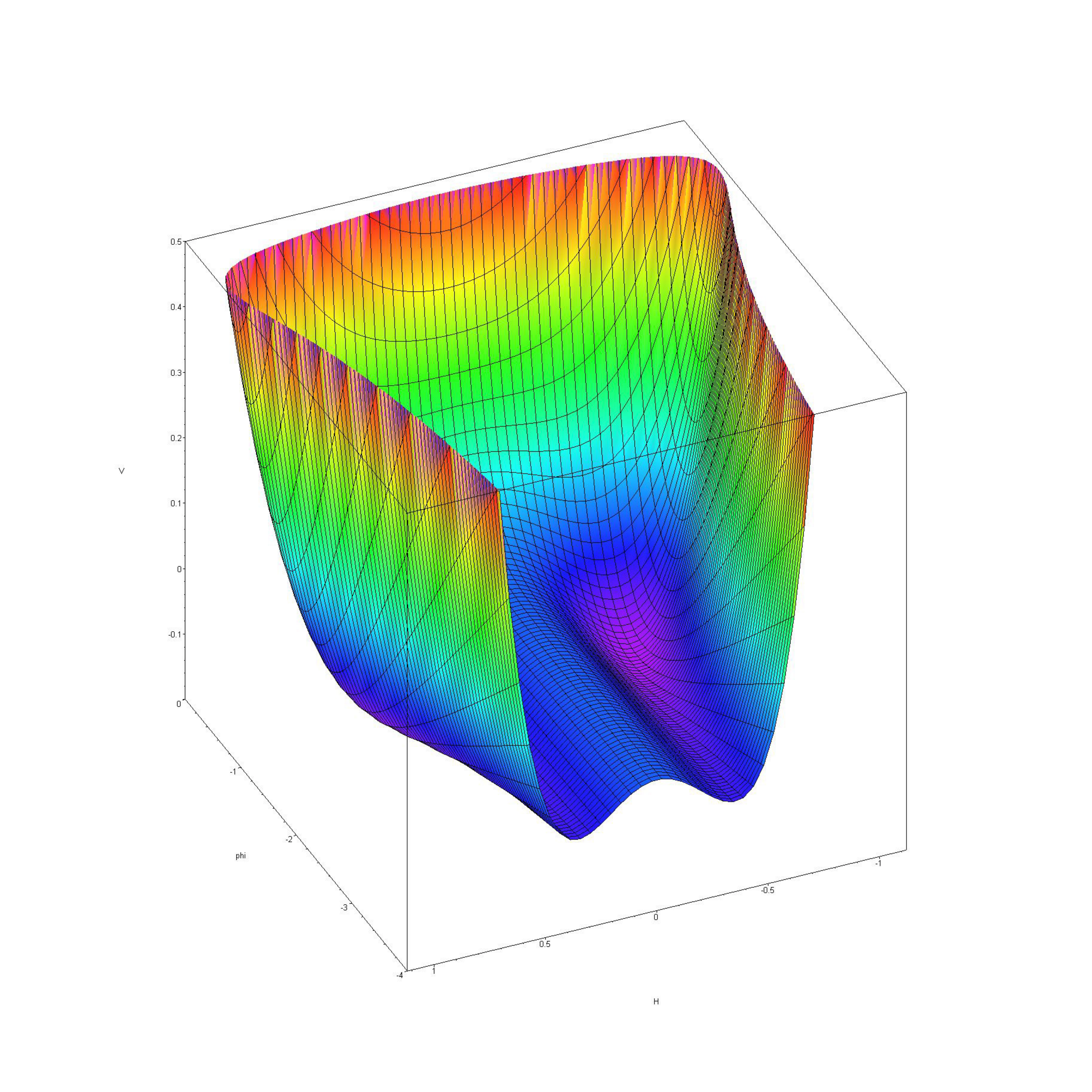} \centering \vskip -0.5cm \caption{The
effective Higgs-dilaton potential in the vicinity of its two
symmetric local minimums: $H^2 =  H^2_m = 0.31$ and $\phi = \phi_m
= -1.38$. Black lines represent the sections of the plot of the
potential by the surfaces of constant $\phi$ and constant $H$.
Parameters are taken as follows: $A = 1$, $B = 2.1$, $C = 1$, $E =
-1$.} \label{fig1} \vskip -0.5cm
\end{figure}Evidently the phase transition point during evolution appears for
$\phi_{crit} = (1/2) \ln(- E/B) < \phi_{in}$. It can be shown that
$\phi_{fin} <\phi_{crit} <0$ and therefore $B+E > 0$. We remark that
the latter inequality entails $|E_{min}| >|E|$ . As well in this case for $\phi_{in} \leq 0$ the Higgs potential is bounded below for any value of Higgs fields.

By the way we notice that due to \eqref{B},\eqref{E}, critical point
$\phi_c$ coincides with $\phi_0$  and the latter comes from the
invariant action $\hat W_{\mathrm{inv}}$  \eqref{WINV}.  Thereby the
requirement $\phi_c = \phi_0 <0$ means that the invariant potential
$-C\phi_0 H^4$, corresponding to $\hat W_{\mathrm{inv}}$ is bounded
bellow.

\begin{figure}[htb]
\vskip -1.0cm
\includegraphics
[scale=.2]{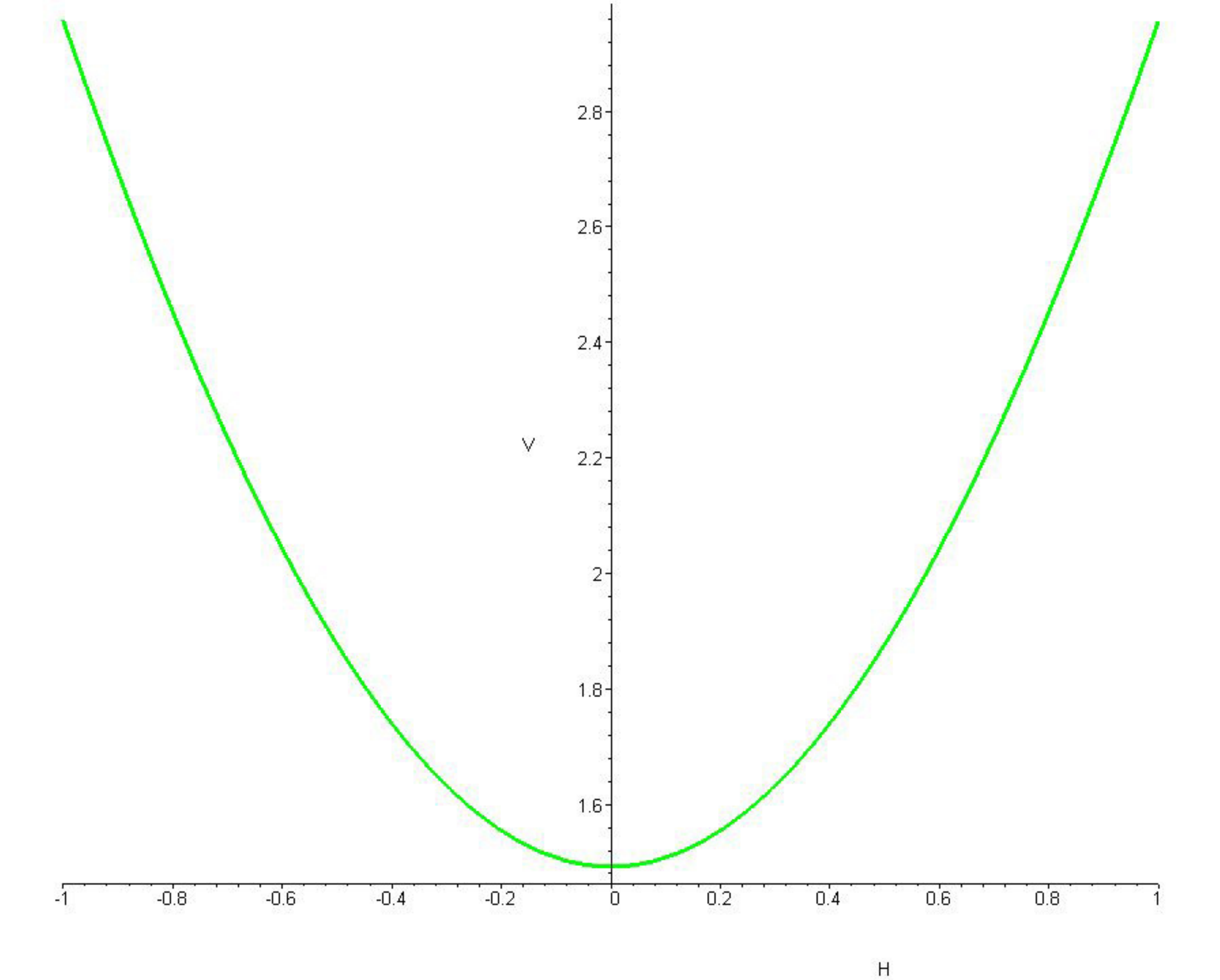}\centering \vskip -0.3cm \caption{$V(H)$ at the
fixed value of $\phi = \phi_{in} = -0.1$ i.e. the profile of the
potential in the symmetric phase. $A = 1$, $B = 2.1$, $C = 1$, $E
= -1$. } \label{fig2}
\end{figure}

We summarize our finding in Fig.~\ref{fig1}. One can see that for
the values of $\phi \simeq 0$ there is only one minimum of the
function $V(H)|_{\phi = fixed}$ at $H = 0$. When we get closer to
$\phi_m$ crossing $\phi_{crit}$, the phase transition occurs, and
every function $V(H)|_{\phi = fixed}$ has two symmetric minimums.
The section of this three-dimensional plot in the initial point
$\phi_{in}=0 $ is shown in Fig.~\ref{fig2} and reveals the absolute
minimum in Higgs fields.

\begin{figure}[htb]
\includegraphics
[scale=.2]{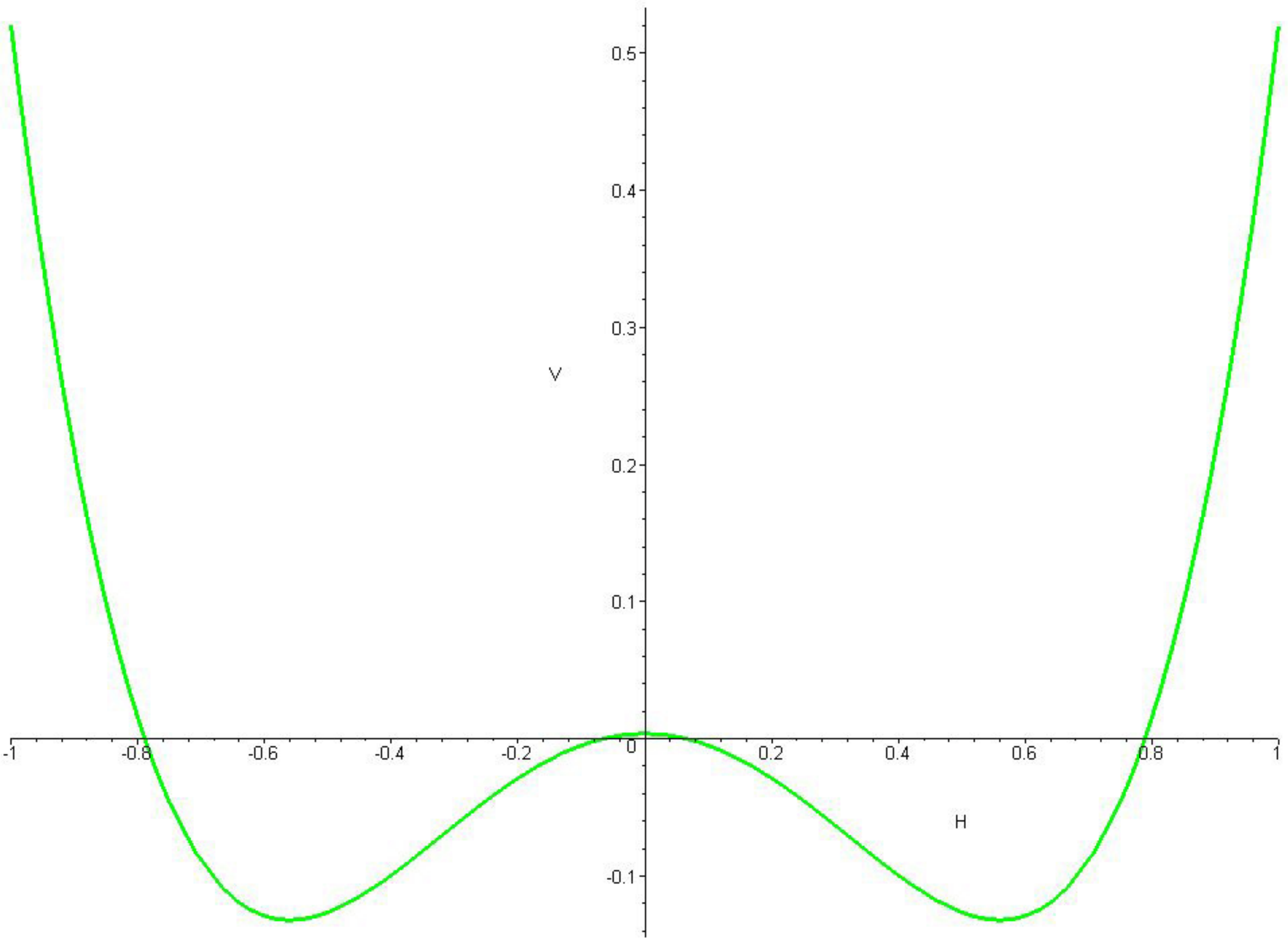}\centering \vskip -0.3cm \caption{$V(H)$ at
the fixed value of $\phi = \phi_m = -1.38$ i.e. the profile of the
potential in the Higgs phase. $A = 1$, $B = 2.1$, $C = 1$, $E =
-1$. }\label{fig3}
\end{figure}

Such a choice of the parameters provides an existence of the local
minimum in the late stage of the universe evolution at $\phi_m =
-1.38$ and $H_m^2 = 0.31$. So in the Higgs phase ($\phi  = \phi_m $)
one has the following potential behavior, $ V(H, \phi_m) = 0.0039 -
0.87H^2 + 1.38H^4$, see Fig.~\ref{fig3}.

In the next Fig.~\ref{fig4}
\begin{figure}[htb]
\includegraphics
[scale=.64]{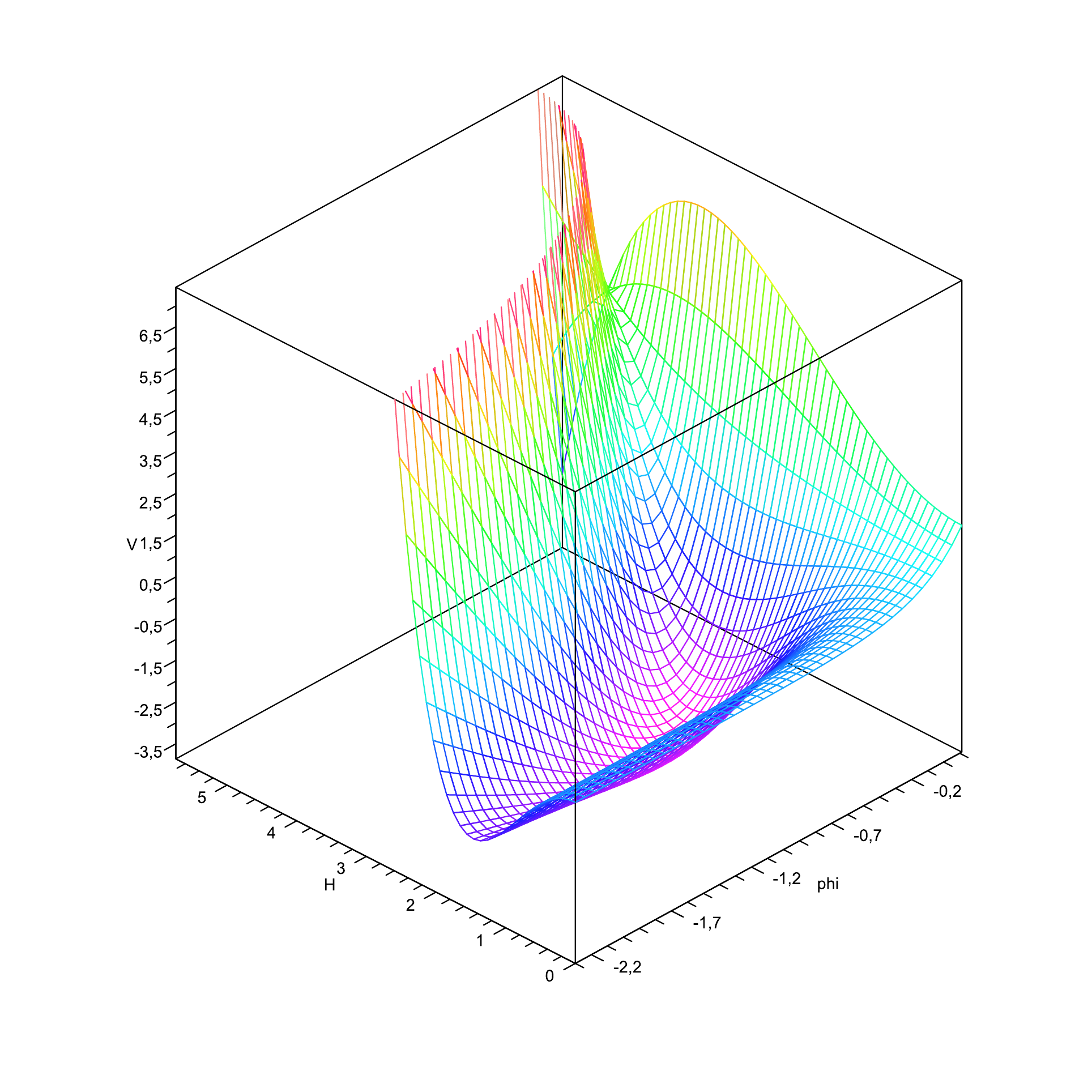}\centering \vskip -0.3cm \caption{The similar
effective Higgs-dilaton potential in the vicinity of its local
minimum: $H =  H_m =  2.29$ and $\phi = \phi_m = -0.72$ chosen to
display  the saddle point. Colored  lines represent the sections
of the plot of the potential by the surfaces of constant $\phi$
and constant $H$.  Parameters are taken as follows: $A = 1$, $B =
2.1$, $C = 0.2$, $E = -2$.} \label{fig4}
\end{figure}
another view on the plot for effective
potential is taken in order to demonstrate that the saddle point is
aside of the steepest descend path.
\subsection{Remark.}
Let us notice, that one is not allowed to identify the minimal value of HD-potential $V_{min}$ with
cosmological constant, because at $A,B,C >0, -B < E< 0$ we can easily prove, that $V_{min}<0$.

Indeed, our potential \eqref{1} satisfies the following relation:
\be
V = \frac{1}{4}\left( \frac{\partial V}{\partial\phi} + H\frac{\partial V}{\partial H} \right)+
\frac{H^2}{4}\left(CH^2 + 2E \right) + V_0\label{relat}
\ee
and thereby its minimal value is given by
\be
V_{m} =V_0 +\frac{H_m^2 }{4}\left(CH^2 + 2E\right). \label{vmin}
\ee
For a given value of the Higgs v.e.v. $H_m\equiv \eta_{fin}e^{2\phi_{fin}}$ one can present the coefficient $E$ in the form
\be
E = H_m^2 \cdot C W_{-1}\left(\frac{E e^{-\frac{B}{C\eta_{fin}}}}{C\eta_{fin}}\right) \la{E1}.
\ee
Substituting \eqref{E1} into \eqref{vmin} we have:
\be
V_m = V_0 + \frac{C H^4}{4}\left(1+2W_{-1}\left(\frac{E e^{-\frac{B}{C\eta_{fin}}}}{C\eta_{fin}}\right)\right)
\ee
and taking into account, that $V_0 <0$, $W_{-1} \leq -1$ we finally obtain:
\be
V_m <  -\frac{CH_m^4}{4} < 0.
\ee
Anyway we suppose, that the observed cosmological constant is generated by both visible and dark matter, and
only visible matter participates in the dilaton - bosonization process considered above and hence
$V_m$ can be identified with (negative) contribution to the total (positive) cosmological constant.


Let's use first the metric $g_{\mu\nu}$ as a background one and therefore independently in the dark and visible sectors,
\[
Z_{\mathrm{total}}(g_{\mu\nu})=Z_{\mathrm{dark}}(g_{\mu\nu})\cdot Z_{\mathrm{SM}}(\tilde g_{\mu\nu}, H)\Big|_{\tilde g_{\mu\nu} = g_{\mu\nu}} .
\]
Performing bosonization $\tilde g_{\mu\nu} \to \tilde g_{\mu\nu}
\exp(2\phi);\ H \to H \exp(-\phi)$ in the SM sector only one
finds,\[ Z_{\mathrm{SM}}(\tilde g_{\mu\nu}, H) = \hat
Z_{\mathrm{inv}}(\tilde g_{\mu\nu}, H) \int d\phi\,
e^{-S_{\mathrm{\mathrm{coll}}}(\tilde g_{\mu\nu}, H, \phi)} \simeq
\int d\phi e^{-\int d^4x \sqrt{- g} V (\tilde g_{\mu\nu}, H, \phi)}
.
\]
The total cosmological generating functional is produced after averaging over gravity, i.e. over the metrics,
\[
Z_{\mathrm{cosm}}= \int {\cal D} g_{\mu\nu} \times [\mathrm{gauge \ fixing}]\times e^{- W_{grav}(g)} Z_{\mathrm{dark}}(g_{\mu\nu})\cdot Z_{\mathrm{SM}}(g_{\mu\nu}, H) .
\]
The latter integral in the vacuum energy approximation for matter fields entails the determination of the cosmological constant,
\[Z_{\mathrm{cosm}} \sim \int {\cal D} g_{\mu\nu} \times [\mathrm{gauge \ fixing}]\times e^{- W_{grav}(g) - \int d^4x \sqrt{- g} \frac{\Lambda_{cosm}}{8\pi G_N}}, \]
which evidently consists of,
\[\frac{\Lambda_{cosm}}{8\pi G_N} = V_{0,SM} +V_{0,\mathrm{dark}}, \quad  \int d^4x \sqrt{- g} V_{0,\mathrm{dark}}  \simeq - \log Z_{\mathrm{dark}} . \]
All formulas are referred to the Euclidean space-time and can be re-written easily for the Minkowski one.

\section{Conclusions}
In this paper we have seen how the bosonic spectral action emerges
form the fermionic action and Weyl anomaly via the renormalization
group flow. In this sense we can say that the bosonic degrees of
freedom are induced by the fermionic ones. The procedure followed is
spectral and therefore well suited for the noncommutative approach
to the standard model. The action emerges, in case of the presence
of a fundamental scale, and therefore of a non Weyl invariant
fundamental theory, in terms of a composite dilaton.

What we find particularly encouraging is the fact that, at the level
of effective potential, the theory gives rise to a Higgs-dilaton
potential with desirable qualitative features, i.e.\ the presence of
both a broken and an unbroken phase, and the possibility to roll
form the latter to the former. We did so using just the bare
essential ingredients of the spectral action, and therefore the
result, while necessarily generic and qualitative, are to a large
extend independent on the details of the model.  We see a certain relevance of the Higgs-dilaton
potential of our type for realization  of the Higgs field assisted inflation and further stages of the Universe evolution undertaken in \cite{shaposh,Barvinsky:2009jd}. A refinement of this
work taking other degrees of freedom into account is possible, and
partially under way.

\acknowledgments This work has been supported in part by CUR
Generalitat de Catalunya under project 2009SGR502, by project FPA2010-20807-C02-01 and by Consolider CPAN. The work of
A.A.A.\ and M.A.K.\ was supported by Grant RFBR 09-02-00073, 11-01-12103-ofi-m and
SPbSU grant 11.0.64.2010. M.A.K.\ is supported by Dynasty Foundation
stipend.

\setcounter{section}{0}
\appendix{Appendix\label{appendix}}

In this appendix we show why $W_{\mathrm{inv}}$ is proportional to
$H^4$. In the approximations taken, when all the fields do not
depend on the space-time coordinates one may infer (up to the
additional terms, proportional to $H^4$) the dependence on the Higgs
field of the initial effective potential $W$, related to $Z$
\eqref{zinitial}, based on the requirement, that $Z_{\mathrm{inv}}$
\eqref{caseA}  is invariant under the Weyl transformation
\eqref{scaleinvariance}.

Since we are interested in the form of the effective Higgs potential
and ignore all derivative terms, one can write $Z(H)$ instead of
$Z(D)$, and consider ordinary integrals over $\phi$ instead of
functional.

We have seen~\eqref{zinvprod}, that the invariant partition function
can be rewritten as:
\be
Z_{\mathrm{inv}}(H) = Z(H)\cdot Z_{\mathrm{anom}}(H) =
e^{-W(H)}\cdot \int d\phi e^{-S_{\mathrm{anom}}(\phi,H)},\la{zinv}
\ee
where $S_{\mathrm{anom}}\equiv \int d^4 x \sqrt{|g|}
V_{\mathrm{anom}}$ and due to \eqref{sanomcoll} and \eqref{vcoll}:
\ba V_{\mathrm{anom}} = -V_{\mathrm{coll}}=-\int d^4 x \sqrt{|g|}
\left\{ \tilde A\left(e^{4\phi} - 1\right) + \tilde B H^2\left(
e^{2\phi} -1\right) - C\phi H^4 \right\} \la{sanom}, \ea Where \ba
\tilde A &=& \frac{45}{8\pi^2}\frac{\Lambda^4}{8}\left(2\log\frac{\Lambda^2}{\mu^2}-1\right),\la{A}\\
\tilde B &=& \frac{3 y^2}{2\pi^2}\frac{\Lambda^2}{2}\left(1 - \log\frac{\Lambda^2}{\mu^2}\right),\la{B} \\
C &=& \frac{3 z^2}{4\pi^2}.\la{C}
\ea

Notice, that we can perform integration over $\phi$ in \ref{zinv}
via the Laplace method. Indeed, in our approximation we neglect all
the (covariant) derivatives, thats why $S_{\mathrm{anom}}$ is
proportional to the volume $vol$ of space-time (we have an infrared
cutoff implicit in the theory).

\ba \int d\phi e^{-(vol)\cdot V_{\mathrm{anom}}(H,\phi)}&\simeq&
\sqrt{\frac{2\pi}{vol\cdot
\frac{\partial^2V_{\mathrm{anom}}(H,\phi_{min})}{\partial\phi^2}}}e^{-(vol)\cdot
V_{\mathrm{anom}}(H,\phi_{min})} \no &=& e^{-(vol)\cdot
\left(V_{\mathrm{anom}}(H,\phi_{min}) +
\frac{O(\ln{(vol)})}{vol}\right)}. \ea Taking into account, that
$vol$ goes to infinity we must take into account only the leading
term in the exponent. Thereby one can finally write:
\be
W_{\mathrm{anom}}(H) \equiv e^{-W_{\mathrm{anom}}(H)}, ~~~
W_{\mathrm{anom}} \equiv {\int d^4 x
\sqrt{|g|}V_{\mathrm{anom}}(H,\phi=\phi_{min}(H))}.\la{wanom2}
\ee

For a fixed value of $H$ the function
$V_{\mathrm{anom}}(\phi,H=fixed)$ has a local minimum at the point:
\be
\phi_{min} (H) = \frac{1}{2} \ln  \left( {\frac { \left( \tilde B+\sqrt {{\tilde B}^{2}+4\,
\tilde A C} \right)  H^2}{-4\tilde A}} \right) \la{phim},
\ee
if and only if $\tilde A < 0,~\tilde B > 0$. Thereby we have the
following expression for  $V_{\mathrm{anom}}$:
\be
V_{\mathrm{anom}} = \tilde A + \tilde B H^2 +  \frac{C H^2}{2} \ln
\left( {\frac { \left( \tilde B+\sqrt {{\tilde B}^{2}+4\, \tilde A
C} \right)  H^2}{-4\tilde A}} \right) + H^4 \cdot
const(A,B,C).\la{Vanom}
\ee

By the construction we know, that $Z_{\mathrm{inv}}$ is unchanged
under the transformation \ref{scaleinvariance}. It means, that
$W_{\mathrm{inv}}(H) = W(H) + W_{inv)(H)}$ must be invariant under
the transformation
\be
g_{\mu\nu}\rightarrow e^{2\phi}g_{\mu\nu},\no
H \rightarrow e^{-\phi}H.
\ee
Thereby we have the following functional equation for $V(H)$ \ba
V(e^{-\phi} H) + V_{\mathrm{anom}}(e^{-\phi} H)  = e^{-4\phi}(V(H) +
V_{\mathrm{anom}}(H)).\la{feq} \ea One can easily see, that the most
general solution of the \ref{feq} is given by\footnote{The way of
solving of \eqref{feq} looks as follows. Every solution of \ref{feq}
satisfies the differential equation, obtained from \eqref{feq} by
differentiating of both left- and right-hand sides of it over $\phi$
and substituting $\phi = 0$. The general solution of the latter is
given by \eqref{V}. One can check, that \eqref{V} satisfies to
\eqref{feq}, so we conclude, that we have found the general solution
of \eqref{feq}.}
\be
V = -\tilde A  -\tilde B H^2   -\frac{C H^2}{2} \ln  \left( \frac{H^2}{\tilde\mu} \right) ,\la{V}
\ee
where $\tilde\mu$ is an integration (dimensionful) constant. Finally
we obtain from \eqref{V} and \eqref{Vanom} (as we expected):
\be
V_{\mathrm{inv}} = V + V_{\mathrm{anom}} = \tilde\gamma H^4,
\ee
with some (undefined at this stage) constant $\gamma$. We emphasize,
that the potential $V$ given by \eqref{V} has a local minimum at
$\tilde A <0, \tilde B > 0$, but after averaging over dilatations
the minimum disappears.

We also notice, that we can perform similar integration via the
Laplace method in the \eqref{Zscen2}. In this case minimum of
$V_{\mathrm{coll}}  = -V_{\mathrm{anom}}$ is of our interest, i.e.
maximum of $V_{\mathrm{anom}}$. The latter is given by
\be
\phi_{max} (H) = \frac{1}{2} \ln  \left( {\frac { \left( \tilde
B-\sqrt {{\tilde B}^{2}+4\, \tilde A C} \right)  H^2}{-4\tilde A}}
\right) ,
\ee
and it exists both for positive and negative signs of $\tilde A$,
and $\tilde B>0$. In this case one has the same expression \eqref{V}
for $V$, which is defined, we remind, up to the invariant term,
proportional to $H^4$. In this sense the point of view taken in this
paper, that is of a partition function which is not Weyl invariant,
and that of reference~\cite{anlizzi}, are not in fundamental
contradiction..

\end{document}